\documentclass[pra,twocolumn,preprintnumbers,amsmath,amssymb,floatfix,superscriptaddress,showpacs]{revtex4}
\usepackage{graphicx}
\usepackage{graphics}
\usepackage{psfrag}
\usepackage{hyperref}
\usepackage{multirow}
\usepackage{color}

\newcommand{\vphi}{\varphi}

\newcommand{\rmi}{{\rm i}}
\newcommand{\cZ}{c_{{\scriptscriptstyle Z}}}
\newcommand{\cA}{c_{{\scriptscriptstyle A}}}

\begin{document}

\hypersetup{pdftitle={Error estimates and specification parameters for functional renormalization}}
\title{Error estimates and specification parameters for functional renormalization}
\date{\today}
\author{David Schnoerr}
\affiliation{Institute for Theoretical Physics, University of Heidelberg, D-69120 Heidelberg, Germany}
\author{Igor Boettcher}
\affiliation{Institute for Theoretical Physics, University of Heidelberg, D-69120 Heidelberg, Germany}
\author{Jan M. Pawlowski}
\affiliation{Institute for Theoretical Physics, University of Heidelberg, D-69120 Heidelberg, Germany}
\affiliation{ExtreMe Matter Institute EMMI, GSI Helmholtzzentrum f\"{u}r Schwerionenforschung mbH, D-64291 Darmstadt, Germany}
\author{Christof Wetterich}
\affiliation{Institute for Theoretical Physics, University of Heidelberg, D-69120 Heidelberg, Germany}

\begin{abstract}
We present a strategy for estimating the error of truncated functional flow equations. While the basic functional renormalization group equation is exact, approximated solutions by means of truncations do not only depend on the choice of the retained information, but also on the precise definition of the truncation. Therefore, results depend on specification parameters that can be used to quantify the error of a given truncation. We demonstrate this for the BCS-BEC crossover in ultracold atoms. Within a simple truncation the precise definition of the frequency dependence of the truncated propagator affects the results, indicating a shortcoming of the choice of a frequency independent cutoff function.
\end{abstract}

\pacs{05.10.Cc, 11.10.Hi, 67.85.Lm}

\maketitle

\section{Introduction}
Understanding the behavior of strongly correlated quantum many-body systems is an important and challenging task in numerous areas of modern physics, ranging from heavy ion collisions and neutron stars to high temperature superconductors and ultracold atomic gases. Due to the absence of a small expansion parameter, non-perturbative methods have to be employed for the description of these systems. The necessity to go beyond perturbative approaches arises generically in many physical situations, including quantum gravity, systems with largely different length scales, or non-perturbatively renormalizable theories.

Typically, providing error estimates for non-perturbative techniques is hard. For instance, the accuracy of numerical Monte Carlo simulations may be measured by the relative change of observables due to enlarging the number of grid points. Similarly, including higher order terms in truncations of Functional Renormalization Group or Schwinger--Dyson equations partially provides a notion of convergence of the corresponding results. However, such improvements are numerically costly and it is desirable to find methods which allow for error estimates within a given approximation scheme.

A promising direction to quantify experimentally the reliability of non-perturbative methods has emerged with the rapidly evolving field of ultracold quantum gases. The latter can be realized and controlled in experiment with unprecedented accuracy. Observables such as binding energies, the equation of state, the phase diagram or collective modes can be measured over a wide range of interaction strengths, temperatures and densities \cite{Salomon, Navon07052010, Ku03022012}. The precise knowledge of the microscopic Hamiltonian of ultracold alkali gases, combined with the  universality of long range physics, allows for a comparison with first principle methods for computing the partition function which results from this Hamiltonian, and which determines the macroscopic physics of the system. This possibility for precise measurements makes a reliable estimate of errors for non-perturbative methods even more urgent.

In this work, we address the estimate of errors for the functional renormalization group for the effective average action. The simple one loop form of the exact functional flow equation  \cite{Wetterich1993}, together with physically intuitive non-perturbative truncation schemes, makes the method particularly suitable for a study of non-perturbative problems \cite{Berges:2000ew, Pawlowski20072831, Gies:2006wv, Schaefer:2006sr, Delamotte:2007pf, Kopietz2010,  Metzner:2011cw, Braun:2011pp}. In order to have a practical example we discuss possibilities for an error estimate for the use of functional renormalization for the BCS-BEC crossover of non-relativistic fermions in three dimensions. The system is archetypical for the transformation of physical laws when going from the simplicity of the microscopic theory to the complexity of the effective many-body system: Quantum and thermal fluctuations wash out many details of the microscopic laws and new structures such as bound states or collective phenomena emerge. In particular, for infinite s-wave scattering length between the atoms the so-called unitary Fermi gas constitutes a universal strongly coupled system. Important benchmark observables for the latter at zero temperature are the Bertsch parameter, the gap parameter and the Tan contact. Their precise determination attracts a lot of interest, both experimentally and theoretically, and allows for comparison of different non-perturbative approaches in the above sense. Our aim is to obtain an error estimate for the results of the FRG approach to the BCS-BEC crossover for the truncation presented in Refs.~\cite{PhysRevA.76.021602, PhysRevA.76.053627, ANDP:ANDP201010458, Scherer:2010sv, Boettcher:2012cm}. 

This paper is organized as follows. In Sec.~\ref{sec_specification_parameters} we suggest a method to estimate the error of solutions to truncated flow equations by means of specification parameters. We introduce the microscopic model of ultracold fermions in the BCS-BEC crossover and discuss the used truncation scheme in Sec.~\ref{sec_microscopic_model}. In Sec.~\ref{sec_finite_diff} we define specification parameters by means of finite difference projections for the wave function renormalization and gradient coefficient of the inverse boson propagator. In Sec.~\ref{sec_results} we investigate the behaviour of the equation of state, the Bertsch parameter, the gap parameter and the Tan contact upon variations of the specification parameters. In Sec.~\ref{sec_conclusions} we draw conlcusions about the accuracy of the used truncation scheme and discuss how this suggests possibe improvements of the latter.
App.~\ref{append_truncation} contains the truncation in more detail. The flow equations for the running couplings are presented in App.~\ref{append_flow_equations}. The emergence of universality due to fixed point solutions of the running couplings and their corresponding initial values are discussed in App.~\ref{universality}.

\section{Error estimates for functional renormalization}\label{sec_specification_parameters}

The evolution of the effective average action $\Gamma_k$ with the renormalization scale $k$ is governed by an exact flow equation \cite{Wetterich1993}
\begin{equation}\label{flow_equation}\begin{split}
  \partial_k \Gamma_k 
  & = 
    \frac{1}{2}\text{STr} \left[ (\Gamma_k^{(2)} + R_k)^{-1}\partial_k R_k  \right].
\end{split}\end{equation}
Here, $\Gamma_k[\varphi]$ is a functional of bosonic and/or fermionic fields $\varphi$, and $\Gamma_k^{(2)}[\varphi]$ is the second functional derivative or inverse propagator in the presence of arbitrary fields. The supertrace STr includes a single momentum integral, as characteristic for a one loop extremum. The infrared cutoff $R_k$ should guarantee that only a small momentum range of $q^2$ around $k^2$ contributes to Eq.~\eqref{flow_equation}, and that the r.h.s. is ultraviolet and infrared finite.

Despite its simple structure, Eq.~\eqref{flow_equation} is a complicated non-linear functional differential equation. A general functional $\Gamma_k[\varphi]$ involves infinitely many parameters or couplings. Truncations reduce this infinite set to a finite set of couplings or functions that we may call the flowing data $g$. Eq.~\eqref{flow_equation} then translates to a set of flow equations given by
\begin{align}
 \label{Int-1} k \partial_k g = \zeta(g).
\end{align}
Here, $\zeta(g)$ are the beta functions or flow generators for the flowing data $g$. Typically, $g$ contains the effective potential, the propagators and several vertices of the theory. To obtain a numerically tractable set of equations, $g$ can contain at most a finite number of functions. These can be parametrized in different ways, e.g. by a polynomial expansion or function values at given arguments. The approximation procedure for functional renormalization consists in a truncation which specifies the used flowing data $g$, a computation of the corresponding flow generators $\zeta(g)$, and finally a numerical solution of Eq.~\eqref{flow_equation} with \emph{initial data} $g(\Lambda)$ given at some microscopic scale $\Lambda$. For $k \rightarrow 0$ the infrared cutoff $R_k$ is removed, all fluctuations are taken into account, and the data $g(k \rightarrow 0)$ correspond to physical n-point functions or similar quantities. Benchmark quantities as the Bertsch parameter can be extracted from $g(k \rightarrow 0$).

For a given truncated flowing data $g$, the flow generators are not uniquely determined. Indeed, the reduction of the exact flow equation to a finite set of equations introduces the need for a \emph{specification} how $\zeta(g)$ is determined in terms of $g$. This prescription fixes the treatment of couplings which are left out in the flowing data. For example, the inverse propagator of a classical statistical model is a function $P(q^2)$ of the squared  momentum $q^2$ (This holds if the system is invariant under translations and rotations.). One may choose to approximate the momentum dependence of $P(q^2)$ by one simple coupling, $P(q^2)-P(0)=z_k q^2$. The \emph{wave function renormalization} $z_k$ depends on $k$ and is part of the flowing data. One has to specify, however, how one defines $z_k$ precisely. An example is 
\begin{align}
  z_k =
    \frac{P(c^2 k^2)-P(0)}{c^2 k^2}.
\end{align}
(For $c \rightarrow 0$ this amounts to $z_k= \partial P / \partial q^2 |_{q^2=0}$.) The flow generator for $z_k$ (given by $-\eta z_k$, with $\eta$ the anomalous dimension) depends on the choice of the \emph{specification parameter} $c$. A more precise form of Eq.~\eqref{flow_equation} is  then

\begin{align}
 \label{Int-2} k \partial_k g =\zeta(g,c),
\end{align}
where $c$ is a set of specification parameters. 

Besides the specification parameters $c$, the flow generators also depend on further numbers like cutoff parameters, bosonization parameters etc. We may collect them in a set of \emph{flow parameters} $w$ which replace $c$ in Eq.~\eqref{Int-2}. Physical results cannot depend on the choice of parameters $w$. The cutoff is only a technical device and its choice has to drop out in the final results \cite{Berges:2000ew}. The same holds for the procedure of bosonization \cite{PhysRevD.68.025020}, for example by the precise implementation \cite{PhysRevB.62.15471, Baier2005144} of a Hubbard-Stratonovich transformation \cite{PhysRevLett.3.77, Stratonovich} or similar concepts for flowing bosonization \cite{PhysRevD.65.065001}. The specification parameters $c$ are obviously only needed for a given approximation scheme and have no meaning for physical results. 

In a given truncation, however, the results will depend on the choice of $w$. The variation of results within a reasonable range of $w$  therefore provides a simple error estimate for the short-coming of a given truncation. In addition, stability or fragility of the flow equations with respect to changes of $w$ can indicate possible ways to improve the truncation scheme. 

Investigations of the dependence of results on cutoff parameters have been performed in the past by employing a family of cutoff functions \cite{Litim200092}. This had led to criteria for optimized cutoffs \cite{Litim200092,PhysRevD.64.105007, Pawlowski20072831, PhysRevD.76.105001, PhysRevD.83.085009}. Also the dependence on free parameters in partial bosonization has been studied \cite{PhysRevD.68.025020}, showing how the ``Fierz Ambiguity'' in the Hubbard-Stratonovich transformation is resolved. The influence of the relative cutoff scale on observables in a theory of both bosons and fermions has been investigated in \cite{PhysRevA.81.043628, PhysRevA.83.023621, PhysRevC.78.034001}. 

In this paper we concentrate on the influence of the specification parameter $c$ on the FRG approach to the BCS-BEC crossover in the truncation of Refs.~\cite{PhysRevA.76.021602, PhysRevA.76.053627, ANDP:ANDP201010458}. For this quantum statistical system the inverse boson propagator $P_{\phi}(q_0, q^2)$ is a function of two variables, the squared space-momentum $q^2$ and the time component $q_0$ that is related to the Matsubara frequency. The truncation of $\Gamma_k$ uses for each $k$ a simple form for the inverse propagator,
\begin{align}
 \label{Int-3} P_{\phi,k}(q_0,q^2) = \rmi Z_k q_0 + \frac{1}{2} A_k q^2,
\end{align}
with flowing data comprising the two ``couplings''  $Z_k$ and $A_k$. In Refs.~\cite{PhysRevA.76.021602, PhysRevA.76.053627, ANDP:ANDP201010458} the corresponding flow equations for the running couplings $Z_k$ and $A_k$ have been defined by the derivative of the flow equation for the boson propagator at vanishing momentum and vanishing frequency. In this work, we generalize this prescription by means of a finite difference projection according to
\begin{align}
 \label{Int-4} \partial_k Z_k = \frac{\partial_k P_{\phi,k}(c_{\scriptscriptstyle Z}^2 k^2,0) - \partial_k P_{\phi,k}(0,0)}{\rmi c_{\scriptscriptstyle Z}^2 k^2},
\end{align}
and similarly for $A_k$ with a parameter $\cA$. Although the flowing data $g=(Z,A)$ remains the same, the flow generators additionally depend on the specification parameters $\cZ$ and $\cA$.  Indeed, an exact flow equation for $P_{\phi,k}(q_0,q^2)$ can be derived from the second functional derivative of Eq.~\eqref{flow_equation} \cite{Berges:2000ew}. Even for a given truncation of the vertices appearing in this equation the flow generators for $Z_k$ and $A_k$ will depend on the point in momentum space where $\partial_k P_{\phi,k}$ is evaluated, and therefore on the choice of $\cZ$ and $\cA$.

The motivation for introducing the finite difference projection in Eq.~\eqref{Int-4} resides in the fact that, for a proper choice of regulator functions, the loop integral appearing in the flow generators at a given scale $k$ is dominated by values of the propagator with $q_0 \approx k^2$ and $q^2 \approx k^2$. Hence, the truncation should resolve the inverse propagator in the vicinity of these in a sufficient manner. This is ensured by the finite difference projection with $\cZ$ and $\cA$ near unity. In many situations a continuation to $\cZ, \cA \rightarrow 0$ induces only a small effect, while values of $\cZ$ and $\cA$ substantially larger than unity are not meaningful. 

In the FRG approach to the BCS-BEC crossover, so far, only regulator functions which do not depend on frequencies have been employed for the sake of technical simplicity. Unfortunately, this implies that both very large and very small frequencies (as compared to $k^2$) contribute to the loop integrals in the flow generators. Only the momentum integrals over space-momenta $\vec{q}$ are effectively restricted to a small interval. Because of this shortcoming in the choice of $R_k$ one of the main advantages of the flow equation \eqref{flow_equation}, namely the effective restriction of all momentum integrals to a small range, is lost. Linear approximations to the $q_0$-dependence of the inverse propagator $P_{\phi,k} (q_0,q^2)$ may therefore become insufficient. A substantial dependence of results on $\cZ$ could reveal such an insufficiency, and we indeed will find that this is the case. This demonstrates that our approach to an error estimate can be used directly to detect particular shortcomings of a given truncation or choice of cutoff.

To investigate the relevance of $\cZ$ and $\cA$, we compute the equation of state, the gap parameter, the Tan contact and the dimer-dimer scattering length in the crossover as a function of the specification parameters. For simplicity, we restrict to the case of zero temperature. We find that the parameter $\cA$ has no important influence on the results within our numerical resolution. In contrast, the results depend significantly on $\cZ$. A detailed discussion is presented in Sec.~\ref{sec_results} and our Conclusions \ref{sec_conclusions}. This result indicates that the regulator works well for spatial momenta, but fails to effectively cut off the contributions of frequencies outside an interval $q_0 \approx k^2$. As a result the detailed frequency dependence of the boson propagator matters, and a too simple approximation for this frequency dependence leads to substantial inaccuracies of the results. Incorporating a regulator function which depends on both spatial momenta $q^2$ and frequencies $q_0$ may result in observables which are much more insensitive to variations in both $\cZ$ and $\cA$.

\section{Microscopic model and Truncation}\label{sec_microscopic_model}
We consider an ultracold Fermi gas in three dimensions, where the atoms can occupy two different hyperfine states. We assume a balanced population. The atoms are described by a two-component Grassmann field $\psi=(\psi_1, \psi_2)$. The collective bosonic degrees of freedom are incorporated in a complex scalar field $\varphi$. The euclidean microscopic action is given by
\begin{equation}\begin{split}
  S[\psi, \varphi] = 
  & 
    \int_0^{1/T} \text{d} \tau \int \text{d}^3 x \Big( \psi^{\dagger} \left( \partial_{\tau} - \nabla^2 - \mu \right) \psi \\
  & \quad
	+  \varphi^* (  \nu_{\Lambda} - 2\mu ) \varphi -  h \left( \varphi^*\psi_1 \psi_2 - \varphi \psi_1^* \psi_2^* \right) \Big).
\end{split}\end{equation}
It is fixed at a sufficiently large momentum scale $\Lambda$, where the interactions between the atoms can be approximated to be pointlike. Hence, $\Lambda^{-1}$ has to be much larger than the characteristic length scale where details of the interatomic potential are resolved. The latter is of the order of the van-der-Waals length or the effective range. 

We work in non-relativistic natural units with $\hbar = k_{\rm B} = 1$ and rescale $2M=1$, where $M$ is the mass of the fermionic atoms. The fields depend on the euclidean time $\tau$, which is restricted to a torus with circumference $1/T$, where $T$ is the temperature of the system. We introduce the Feshbach or Yukawa coupling $h$, which is directly related to the width of the Feshbach resonance \cite{PhysRevA.73.033615, Diehl2007206}. After renormalization, the term $\nu_{\Lambda}$ corresponds to the detuning from resonance. This parameter can be varied experimentally by applying an external magnetic field.

The basic idea underlying the Functional Renormalization Group is to take the microscopic action $S$ at ultraviolet scale $\Lambda$ as a starting point and to successively include quantum and thermal fluctuations of momenta larger than a flowing momentum scale $k$. For $k=0$, all fluctuations are included and one arrives at the full effective action $\Gamma$. The latter, in turn, is the generating functional of the one-particle irreducible correlation functions of the theory, and thus contains all information about the macroscopic system. At intermediate values of $k$, the effective average action $\Gamma_k$ can be interpreted as the effective action for a theory valid at momentum scale $k$.  In summary, the functional $\Gamma_k$ satisfies $\Gamma_{\Lambda}=S$ and $\Gamma_0 = \Gamma$. Its evolution is determined by the exact flow equation \eqref{flow_equation}. More precisely, in this equation the ``supertrace'' \text{STr} denotes an integration over momenta and a summation over field indices, with the characteristic minus sign for fermions. The regulator $R_k$ has to satisfy
\begin{equation}\begin{split}
  \lim_{q^2/k^2 \rightarrow 0} R_k(Q) & = k^2, \\
  \lim_{q^2/k^2 \rightarrow \infty} R_k(Q) & = 0, \\
\end{split}\end{equation}
with $Q=(q_0, \vec{q})$. In principle, one would like to have a similar property for the dependence of $R_k$ on $q_0$, but this is not realized for the cutoff functions depending only on $q^2$ that have been employed so far.

In this work, we use a basic and physically motivated truncation for the effective average action of the BCS-BEC crossover. The momentum dependence of the inverse boson propagator is assumed to be of the form
\begin{equation}\label{ansatz_boson_propagator}\begin{split}
  P_{\phi,k}(Q)= \text{i} Z_k q_0 + A_k q^2/2.
\end{split}\end{equation}
This ansatz constitutes a systematic infrared expansion of the most general form of the bosonic self-energy, see Eq.~\eqref{appA-3}. A Taylor expansion around the point $(q_0,q^2)=(\cZ^2 k^2, \cA^2 k^2)$ would lead to corrections $\sim (q_0 - \cZ^2 k^2)^2$ and $(q^2 - \cA^2 k^2)^2$ that we neglect here. This is well motivated if the coefficients of these corrections are not too large and if the effective momentum range contributing to the flow is restricted to the neighborhood of $k^2$. In contrast, if the effective momentum range is large, as for $q_0$ in case of an $q_0$-independent cutoff, the deviations of the true propagator from the ansatz \eqref{ansatz_boson_propagator} may have important effects.

We neglect fluctuation effects on the fermion propagator and the Feshbach coupling. The running couplings $Z_k$ and $A_k$, respectively, are referred to as wave function renormalization and gradient coefficient in what follows. We include a scale-dependent effective potential $U_k(\rho, \mu)$, which is a function of the $U(1)$-invariant $\rho = \phi^* \phi$ and the chemical potential $\mu$. It is expanded around its scale-dependent minimum $\rho_0=\rho_{0,k}$ and the chemical potential. Our truncation is discussed in more detail in App.~\ref{append_truncation}.

The choice of regulators is a central issue for the solution of the flow equation. To implement the idea of a momentum-shell integration at each renormalization group step, one should use cutoff functions $R_k(Q)$ which are localized at $q^2 \approx k^2$ and $q_0 \approx k^2$ for a given scale $k$. In practical applications of the FRG, however, one often relies on regulators which only depend on spatial momenta $q^2$. This simplifies the computation of the flow generators, since Matsubara summations over frequencies can be carried out analytically. Moreover, this choice typically  provides qualitatively convincing results. Although optimization criteria for regulators can be derived, it is, at present, not fully understood how much the results of FRG calculations are influenced by the choice of the function $R_k(Q)$. 

In this work, we use optimized regulator functions
\begin{align}\label{boson_regulator_function}
  R_{\phi, k}(Q) 
  & =
    A_k (k^2 - q^2/2)\theta(k^2-q^2/2)
\end{align}
and
\begin{align}
\label{fermion_regulator_function}
  R_{\psi, k}(Q) 
  & = 
    (k^2 \text{sgn}(q^2 - \mu) - (q^2 - \mu))\theta(k^2-\vert q^2 - \mu \vert),
\end{align}
which do not depend on $q_0$. For the bosons, we suppress fluctuations with small momenta, whereas for the fermions we regularize around the Fermi surface.

\section{Finite Difference Projection}\label{sec_finite_diff}
\begin{figure}[t]
  \centering
  \includegraphics[scale= 1.1]{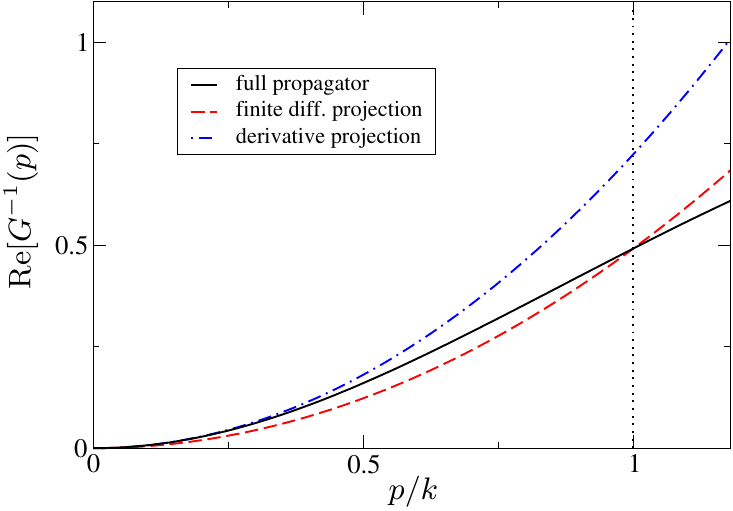}
\caption{Real part of the inverse boson propagator for fixed $p_0$ as a function of momentum for some finite $k$ in arbitrary units. The black solid line represents the full inverse boson propagator, as computed from an integration over the fermion loop. It shows a typical quadratic dependence for small momenta $p/k \ll 1$ and flattening for larger momenta. The blue dot-dashed and red dashed lines show the approximations for the derivative projection and the finite difference projection for $c_{{\scriptscriptstyle A}}=1$, respectively. While the former is appropriate for $p/k \ll 1$, the latter gives a more accurate approximation for $p/k \approx c_{{\scriptscriptstyle A}}$.
}
  \label{fig: finite_diff_motivation}
\end{figure}

In addition to choosing an ansatz for the effective average action, one has to specify projection prescriptions for the scale-dependent parameters. Here, we use established projection prescriptions for the running couplings contained in the effective potential $U_k(\rho, \mu)$, see App. \ref{append_truncation}. The wave function renormalization $Z_k$ and the gradient coefficient $A_k$ are projected by means of finite differences of the inverse boson propagator,
\begin{align}\label{finite_difference_projection_z}
  Z_k &= -\frac{G^{-1}_{\phi,12}(p_0,0)-G^{-1}_{\phi,12}(0,0)}{p_0}\Bigr|_{p_0=c_{{\scriptscriptstyle Z}}^2k^2},\\
\label{finite_difference_projection_a}
  A_k &= 2 \frac{G^{-1}_{\phi,22}(0,p^2)-G^{-1}_{\phi,22}(0,0)}{p^2}\Bigr|_{p=c_{{\scriptscriptstyle A}} k}.
\end{align}
Herein, the inverse propagator matrix $G_{\phi}^{-1}$ is defined as 
\begin{equation}\begin{split}
  G_{\phi,ij}^{-1}(p_0,p^2) \delta(P+P')
  & = 
    \frac{\delta^2 \Gamma_k}{\delta \phi_i(P) \delta \phi_j(P')}.
\end{split}\end{equation}
We work in the real field basis for the bosons with $\phi_1$ and $\phi_2$ given by
\begin{equation}\begin{split}
  \phi = \frac{1}{\sqrt{2}}(\phi_1 + \text{i} \phi_2).
\end{split}\end{equation}
In Eqs.~\eqref{finite_difference_projection_z} and \eqref{finite_difference_projection_a}, we introduce the specification parameters $c_{{\scriptscriptstyle Z}}$ and $c_{{\scriptscriptstyle A}}$ as a measure of the relative width of the corresponding finite difference with respect to the flow parameter $k$. In the limit $c_{{\scriptscriptstyle Z}}, c_{{\scriptscriptstyle A}} \rightarrow 0$, the projections in Eqs.~\eqref{finite_difference_projection_z} and \eqref{finite_difference_projection_a} reduce to derivatives evaluated at vanishing frequency and momentum. This is an established projection prescription used frequently in the literature, in particular for the FRG approach to the BCS-BEC crossover. We refer to it as the ``derivative projection''. While the latter approximates the inverse boson propagator around $p_0=p^2=0$, the finite difference takes into account information at finite frequencies and momenta. It is thus supposed to give more accurate approximations for frequencies and momenta $p_0 \approx k^2$ and $p^2 \approx  k^2$, respectively. Fig.~\ref{fig: finite_diff_motivation} illustrates this idea for the real part of the inverse boson propagator. 

Differentiating both sides of Eqs.~\eqref{finite_difference_projection_z} and \eqref{finite_difference_projection_a}, we find
\begin{align}\label{flow_finite_difference_projection_z}
  \partial_k Z_k &= - \frac{\partial_k  G^{-1}_{\phi,12}(p_0,0)- \partial_k  G^{-1}_{\phi,12}(0,0)}{p_0}\Bigr|_{p_0=\cZ^2k^2},\\
  \label{flow_finite_difference_projection_a}
  \partial_k A_k &= 2 \frac{\partial_k G^{-1}_{\phi,22}(0,p^2)- \partial_k  G^{-1}_{\phi,22}(0,0)}{p^2}\Bigr|_{p=c_{{\scriptscriptstyle A}} k}.
\end{align}
Note that one also has to take into account derivatives of the arguments and the denominator. In App.~\ref{append_flow_equations} we show that these additional terms cancel within the used truncation, however. Due to the bosonic regulator function used in Eq.~\eqref{boson_regulator_function}, only momenta up to $\sqrt{2} k$ contribute to bosonic loops. It thus seems reasonable to choose $c_{{\scriptscriptstyle A}} \lesssim \sqrt{2}$. We consider values of $c_{{\scriptscriptstyle Z}}$ and $c_{{\scriptscriptstyle A}}$ of order unity in the following.

From a technical point of view, the task of solving the flow equations for the scale-dependent parameters of the truncation becomes more involved when introducing the finite difference projections defined in Eqs.~\eqref{flow_finite_difference_projection_z} and \eqref{flow_finite_difference_projection_a}. While all loop integrals can be carried out analytically for the derivative projection, this is not the case for the one entering the flow of $A_k$ with the finite difference projection. A two-dimensional momentum integral remains, which has to be computed numerically at each step during the flow.

\section{Results}\label{sec_results}
By investigating the dependence of observables on $\cZ$ and $\cA$, we are able to estimate the potential error due to a limited momentum and frequency resolution of the inverse boson propagator. To leading order, the bosons are non-dynamical degrees of freedom in the BCS limit. Therefore, we expect that changing the projections for the dynamical components of the inverse boson propagator has only a small influence in this regime. For the unitary and BEC regimes, there is no a priori expectation for the behavior of physical observables.

\begin{figure}[t]
  \centering
  \includegraphics[scale=1.1]{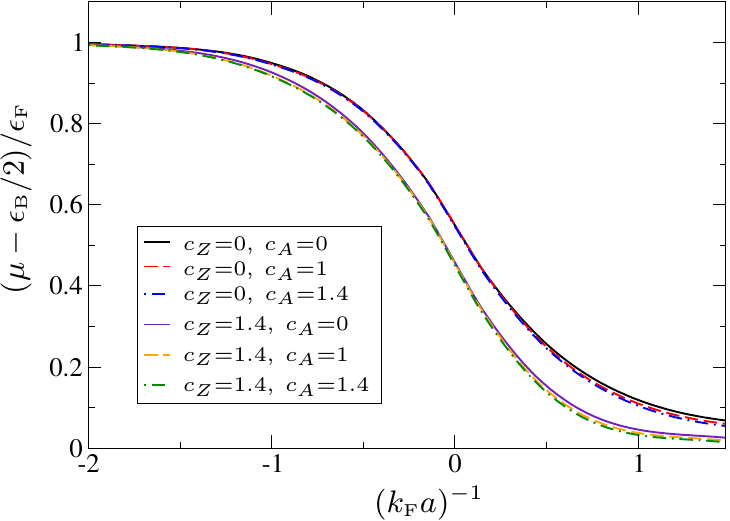}
\caption{The equation of state of the BCS-BEC crossover at zero temperature for specification parameters $\cZ=0$ (upper curves) and $\cZ=1.4$ (lower curves). For each choice of $\cZ$, we vary $\cA$ according to $\cA \in \{0,1,1.4\}$. While the sensitivity of the truncation with respect to $\cA$ is negligible, it significantly depends on $\cZ$. The curve with $\cZ = \cA = 0$ corresponds to the result obtained with the derivative projection in Refs.~\cite{PhysRevA.76.021602, PhysRevA.76.053627, ANDP:ANDP201010458}. 
}
  \label{fig: eos}
\end{figure}

First, we consider the equation of state in the form of the functional dependence of the chemical potential on the density. The black solid line in Fig.~\ref{fig: eos} shows the result of the derivative expansion \cite{PhysRevA.76.021602, PhysRevA.76.053627, ANDP:ANDP201010458, Scherer:2010sv, Boettcher:2012cm}, i.e.~$\cZ=\cA=0$, for $T=0$.
We subtract half the binding energy of a dimer to obtain a positive number. In the BCS regime, we find good agreement with the expected limit $\mu(n) = \epsilon_{\text{F}}(n)$, where $\epsilon_{\text{F}}=k_{\text{F}}^2 = (3\pi^2 n)^{2/3}$ is the Fermi energy of a non-interacting ideal Fermi gas. The derivative expansion also reproduces the Lee--Huang--Yang equation of state of a weakly interacting gas of bosonic molecules in the BEC limit. However, it lacks quantitative accuracy in the unitary regime. For the Bertsch parameter $\xi = \mu / \epsilon_{\text{F}}$ at the unitary point, one obtains  $\xi_\text{d} = 0.55$. (The subscript indicates the used derivative projection.) This value exceeds the experimental finding $\xi_{\text{exp}} = 0.370(5)(8)$ \cite{Ku03022012, 2012arXiv1211.1512Z}. 

The finite difference projection allows us to investigate the sensitivity of the equation of state with respect to variations of $\cZ$ and $\cA$. Fig.~\ref{fig: eos} also shows the equation of state at zero temperature for different values of $\cZ$ and $\cA$. The plot agrees with our expectation that the system does not depend on $\cZ$ and $\cA$ in the BCS regime. Moreover, we observe the result not to depend on $\cA$ for the whole crossover, both for vanishing and nonzero $\cZ$. In contrast, there is a significant decrease of the equation of state for increasing $\cZ$ in the unitary and BEC regimes. The good agreement with Lee--Huang--Yang theory is spoiled for $\cZ>0$. Roughly speaking, for increasing $\cZ$, the result for the equation of state remains unchanged in the BCS regime, improves in the unitary regime, and worsens in the BEC regime.

Table \ref{fig: bertsch_table} lists  the Bertsch parameter obtained from varying $\cZ$ and $\cA$ in the sets $\{0,1,1.4,2.0\}$ and $\{0,1,1.4,2.0,3.0\}$, respectively. We recall that, due to our choice of regulators, spatial momenta are naturally restricted to $q^2 \lesssim 2 k^2$. Therefore, $\cA =1.4$ provides a reasonable upper bound for this specification parameter. We observe a negligible dependence on variations of $\cA$, but a significant decrease for variations of $\cZ$. For $\cZ= 1.4$ we find $\xi(\cZ=1.4) \approx 0.46$ independent of $\cA$.

\begin{figure}[t]
  \centering
  \includegraphics[scale=1.1]{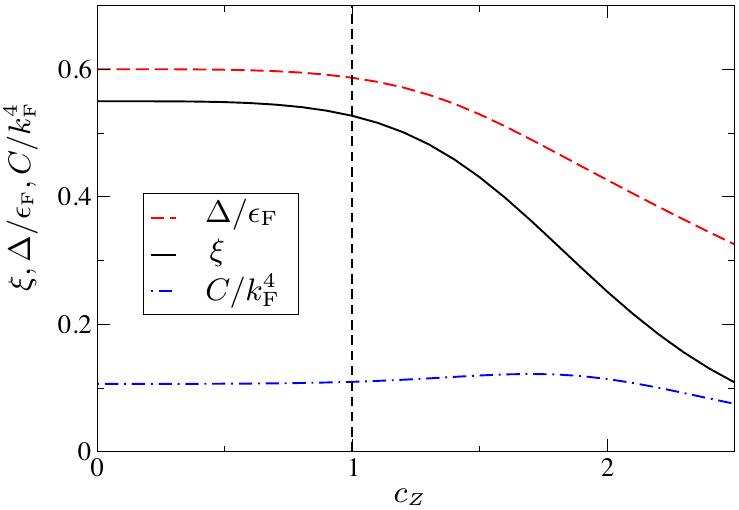}
\caption{Bertsch parameter $\xi$ (black solid line), gap $\Delta/ \epsilon_{\text{F}}$ (red dashed line) and contact $C/ k_{\text{F}}^4$ (blue dot-dashed line) as a function of $c_{{\scriptscriptstyle Z}}$ for fixed $c_{{\scriptscriptstyle A}}=0$ for the unitary Fermi gas. We observe a monotonous behavior of $\xi$ and $\Delta/ \epsilon_{\text{F}}$. The contact $C/ k_{\text{F}}^4$ has a maximum for $c_{{\scriptscriptstyle Z}} \approx 1.7$.
}
  \label{fig: bertsch_gap_c_over_cZ}
\end{figure}

We next discuss variations of $\cZ$ for fixed $\cA=0$. This is motivated by the small dependence of observables  on $\cA$ and the numerical simplicity of choosing $\cA=0$. We compute the Bertsch parameter $\xi$, the gap parameter $\Delta/ \epsilon_{\text{F}}$ and the contact $C/ k_{\text{F}}^4$ for the unitary Fermi gas. Fig.~\ref{fig: bertsch_gap_c_over_cZ} shows the numerical results for these quantities as a function of $\cZ$.  All three quantities show a small dependence for $\cZ\lesssim 1$. They approach the values obtained with the derivative projection in the limit $\cZ \rightarrow 0$. While $\xi$ and $\Delta/ \epsilon_{\text{F}}$ decrease monotonically for $\cZ \gtrsim 1$, the contact $C/ k_{\text{F}}^4$ has a maximum for $\cZ \approx 1.7$ and decreases monotonically for $c_{{\scriptscriptstyle Z}} > 1.7$. The derivative expansion thus constitutes an upper limit for $\xi$ and $\Delta/ \epsilon_{\text{F}}$, whereas the contact only shows slight modifications. We have indicated in Fig.~\ref{fig: bertsch_gap_c_over_cZ} the value $\cZ=1$ which may be considered as a reasonable value for a truncation at a given scale $k$. Comparing the values $\xi=0.55$, $\Delta/ \epsilon_{\text{F}}=0.60$ and $C/ k_{\text{F}}^4=0.11$ at $\cZ=0$ with the values $\xi=0.53~(0.46)$, $\Delta/ \epsilon_{\text{F}}=0.59~(0.55)$ and $C/ k_{\text{F}}^4=0.11~(0.11)$ at $\cZ^2=1~(\cZ^2=2)$ gives an estimate of the error due to the limited frequency resolution of the boson propagator for the used truncation and regularization scheme.

\begin{figure}[t]
  \centering
  \includegraphics[scale=1.1]{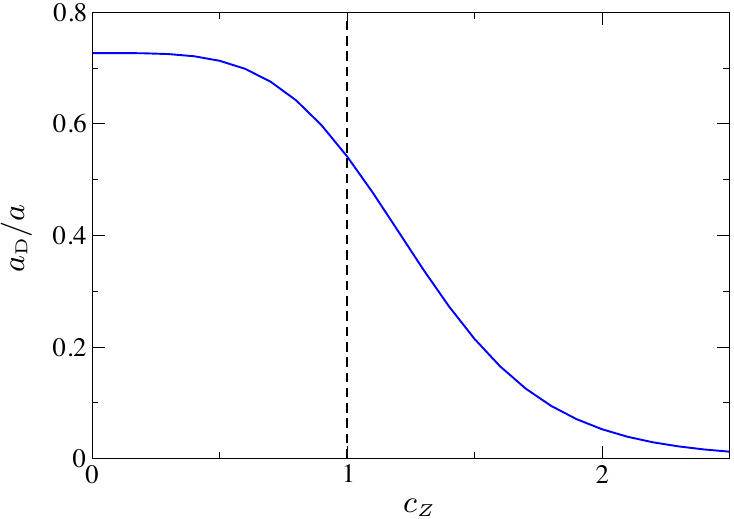}
\caption{The ratio of the dimer-dimer and fermion scattering length $a_{{\scriptscriptstyle\text{D}}}$ and $a$, respectively, as a function of $c_{{\scriptscriptstyle Z}}$ for fixed $c_{{\scriptscriptstyle A}}=0$. The curve monotonically decreases.
}
  \label{fig: scat_ratio}
\end{figure}

Finally, we compute the ratio of the bosonic dimer-dimer scattering length $a_{\text{D}}$ and fermionic atom-atom scattering length $a$. The former is obtained in vacuum in the BEC limit and satisfies $a_{\text{D}} = \lambda_{k=0}/ 4 \pi$ \cite{ANDP:ANDP201010458}. Fig.~\ref{fig: scat_ratio} shows $a_{\text{D}}/a$ as a function of $\cZ$.  We find a monotonous decrease. While the derivative projection results in $a_{\text{D}}/a = 0.73$, the exact value of $a_{\text{D}}/a = 0.6$ \cite{PhysRevLett.93.090404} is matched for $c_{{\scriptscriptstyle Z}} \approx 0.90$. However, we do not expect our truncation to reproduce the correct quantum mechanical value. In order to achieve this, one has to include more vertices in the effective average action, see Ref.~\cite{PhysRevA.81.063619}.

\begin{table}[tbp]
\centering
\begin{tabular}{|l | c c c c|} 
\hline
  $\cA \backslash \cZ$ &  $0$  & $1$  & $1.4$ & $2.0$   
  \\ \hline
  $0$  &  $0.553$  &  $~0.530$  &  $~0.462$  &  $~0.252$ \\
  $1$  &  $0.551$  &  $~0.527$  &  $~0.456$  &  $~0.247$  \\
  $1.4$  &  $0.547$  &  $~0.522$  &  $~0.451$  &  $~0.252$  \\
  $2.0$  &  $0.544$  &  $~0.519$  &  $~0.449$  &  $~0.259$  \\
  $3.0$  &  $0.546$  &  $~0.522$  &  $~0.455$  &  $~0.277$
  \\ \hline
\end{tabular}
\caption{The Bertsch parameter $\xi=\mu/\epsilon_{\rm F}$ for several values of $\cZ$ and $\cA$. Within the numerical uncertainty, the Bertsch parameter is not affected by variations of $c_{{\scriptscriptstyle A}}$. However, it decreases significantly for increasing $c_{{\scriptscriptstyle Z}}$.
}
\label{fig: bertsch_table}
\end{table}

\section{Conclusion}\label{sec_conclusions}

In this work, we have presented a way of quantifying the error of truncated solutions to exact flow equations by means of specification parameters. The latter appear as free parameters in the beta functions of the truncated theory. By varying them within a reasonable range, one can estimate the reliability of a given approximation to the full effective action. A strong dependence of observables on such variations indicates a shortcoming of the truncation, whereas stability hints on a good incorporation of the corresponding physical effect. Both cases have been exemplified here for the BCS-BEC crossover of ultracold two-component fermions.

The interaction between fermionic atoms in our approach is described by the exchange of a bosonic dimer,
\begin{align}
  \lambda_{\psi}
  & = 
    \frac{h^2}{m_{\phi}^2 + P_{\phi}(Q)}.
\end{align}
The propagator of the latter field replaces the effective four-fermion coupling $\lambda_{\psi}$ of the purely fermionic theory. Hence, in order to sufficiently resolve this interaction, a good approximation to the frequency- and momentum-dependence of the boson propagator is mandatory. In a simple truncation, a scale-dependent derivative expansion $P_{\phi,k}(Q) = \rmi Z_k q_0 + A_k q^2/2$, together with a constant part $m_{\phi,k}^2$ evaluated for $Q=0$, comprises the leading terms of a systematic low-energy expansion of the most general form of the boson self-energy. This ansatz for the inverse boson propagator, however, still requires a specification of how the running couplings $Z_k$ and $A_k$ are related to the exact flow equation for the effective average action.

Starting from the exact flow equation for the inverse boson propagator, a possible way of projecting onto the flow of $Z_k$ and $A_k$ consists in taking the derivative with respect to $\rmi q_0$ and $q^2$, respectively, at vanishing external momentum. This procedure has been applied in the literature and results in particular predictions for macroscopic observables like the equation of state or the gap parameter. To obtain an error estimate for the latter quantities, we continuously deform the differentiation by means of a finite difference formula of width $c$ to project onto the couplings $Z_k$ and $A_k$.

More precisely, we define $Z_k$ and $A_k$ by finite differences with different widths, $\cZ$ and $\cA$, respectively. Distinguishing temporal from spatial degrees of freedom is, firstly, motivated from the generic form of the non-relativistic dispersion relation. Secondly, due to our particular choice of cutoff functions which only regulate spatial momenta, the frequency dependence of the boson propagator is expected to be less accurate.

From an investigation of the equation of state, the gap parameter, the Tan contact and the dimer-dimer scattering length in the zero temperature crossover, we find that the specification parameter $\cA$ has no significant influence on the results. The stability of the observables upon variations with respect to $\cA$, which has been verified here for $0 \leq \cA \leq 3$, is not clear a priori. Due to our choice of the regulator, momentum integrals are limited to $q^2 \leq 2 k^2$, thus $\cA$ is limited by $\cA \lesssim 1.4$. For this reason, our upper bound $\cA \leq 3$ provides a solid basis to conclude that the cutoff works well in the spatial domain and that the resolution for spatial momenta is presumably sufficient.

In contrast to the stability upon variations with respect to $\cA$, the specification parameter $\cZ$ substantially modifies the predictions for the above observables. 
This is an indication for the error due to an insufficient frequency resolution of the bosonic propagator. For the cutoff used here the error in the unitary regime is quite substantial, typically $10-20 \%$. A better cutoff function, which also acts as a cutoff for frequencies, may be expected to reduce the error substantially, even within the given simple truncation \eqref{ansatz_boson_propagator}. Thus our example shows how specification parameters can help to obtain an estimate of errors within a given truncation and how to find weak points of an approximation scheme at hand.

We have concentrated here on the specification parameters for the boson propagator. Similar specification parameters should be investigated for the fermion propagator. Also the precise definition of the density matters. Investigating a whole family of possible density definitions will introduce additional specification parameters. In principle, this continues to the specification of vertices. One expects, however, the main uncertainties to arrise from the lowest n-point functions for which truncations are used. The dependence of results on a manageable set of flowing parameters can then be used for an estimate of the error of a given truncation of the flow equations.

\begin{center}
 \textbf{Acknowledgments}
\end{center}
\noindent 
D.~S.~acknowledges financial support from the Studien\-stiftung des deutschen Volkes. I.~B.~acknowledges funding from the Graduate Academy Heidelberg. This work is supported by the Helmholtz Alliance HA216/EMMI and the grant ERC- AdG-290623.

\begin{appendix}

\section{Truncation}\label{append_truncation}
The flow equation for the effective average action constitutes an exact relation. For its practical solution, however, we have to employ a truncation of the most general form of the effective average action, which reduces the infinite number of flowing correlation functions to a finite, computationally manageable set. Several physically motivated choices are possible. Here we use a scale-dependent derivative expansion, where, for each $k$ separately, the functional $\Gamma_k$ is expanded in powers of gradients of the field variables. Typically, the long-range physics are well-captured by this approximation.

Due to fluctuations, the pointlike (thus momentum-independent) interactions of the microscopic theory are replaced by effective vertices with a non-trivial momentum dependence. For the case of the BCS-BEC crossover, an effective four-fermion interaction $\lambda_{\psi,k}(Q_1,Q_2,Q_3)$, a momentum-resolved Feshbach coupling $h_k(Q_1,Q_2)$, and an infinite tower of dimer-dimer and dimer-atom vertices are generated during the flow. In principle, the Functional Renormalization Group allows for incorporating these effects. For the low energy physics, a reasonable choice consists in approximating this complicated vertex structure by pointlike interactions for suitable collective degrees of freedom. 

In this work, we model the atom-atom interactions by an effective boson exchange. The boson can be associated to a dimer or correlated atom pair. Higher order scattering between atoms is partially incorporated in the effective potential, which describes interactions between the dimers. We neglect the regeneration of a four-fermion vertex beyond the dimer exchange and approximate the Feshbach coupling to be given by its microscopic value $h_\Lambda$. For more elaborate truncations of the effective average action for the BCS-BEC crossover, including the four-fermion and atom-dimer vertices, see Refs.~\cite{PhysRevB.78.174528, PhysRevA.81.063619}.

Our ansatz for the effective average action expressed in terms of the mean fields $\bar{\vphi}=\langle \vphi \rangle$ and $\bar{\psi} = \langle \psi \rangle$ reads
\begin{align}
 \nonumber \Gamma[\bar{\vphi},\bar{\psi}] = \int_X &\Bigl( \bar{\psi}^\dagger\bar{P}_{\psi,k}(\partial_\tau,-\rmi \nabla)\bar{\psi} + \bar{\vphi}^* \bar{P}_{\phi,k}(\partial_\tau,-\rmi \nabla)\bar{\vphi}\\
 \label{appA-1} &\mbox{ }-h_\Lambda(\bar{\vphi}^* \bar{\psi}_1\bar{\psi}_2-\bar{\vphi}\bar{\psi}_1^*\bar{\psi}_2^*) + \bar{U}_k(\bar{\rho},\mu)\Bigr).
\end{align}
The effective average potential is a function of the $U(1)$-invariant $\bar{\rho}=\bar{\vphi}^*\bar{\vphi}$. For a given scale $k$, the inverse fermion and boson propagators, respectively, are given by their classical parts contained in the microscopic action, and $k$-dependent self-energy corrections $\Sigma_k(Q)$, which are functions of both frequencies and momenta. Thus we have
\begin{align}
 \label{appA-2} \bar{P}_{\psi,k}(Q) &= \rmi q_0 + q^2 -\mu +\bar{\Sigma}_{\psi,k}(Q),\\
 \label{appA-3} \bar{P}_{\phi,k}(Q) &= \bar{\Sigma}_{\phi,k}(Q).
\end{align}
In addition to $\bar{P}_{\phi,k}$ the inverse propagator for the boson contains a momentum- and frequency-independent part from the second derivative of the effective potential $U_k$.
For the present work, we restrict to a leading order derivative expansion of the boson self-energy, and neglect corrections to the classical fermion propagator. We write
\begin{align}
 \label{appA-4} \bar{P}_{\psi,k}(Q) &= \rmi q_0 + q^2 -\mu,\\
 \label{appA-5} \bar{P}_{\phi,k}(Q) &= \rmi Z_k q_0 + \frac{1}{2} A_k q^2.
\end{align}
We emphasize that the $k$-dependent derivative expansion of the boson propagator in Eq.~\eqref{appA-5} goes beyond the classical quadratic dispersion relation. Indeed, for a sufficiently local regulator $R_{\phi,k}(Q)$, the contributions to the flow of $Z_k$ and $A_k$ are dominated by fluctuations with $q_0 \approx k^2$ and $q \approx k$. For $k \ll |Q|$ the flow of $\bar{P}_{\phi,k}(Q)$ is suppressed by inverse powers of $|Q|$ such that $Z_k$ and $A_k$ in Eq.~\eqref{appA-5} can be replaced by $Z(Q) = Z_{k=|Q|}$ and $A(Q) = A_{k=|Q|}$ in a rough approximation.

Therefore, the ansatz $\rmi Z_k q_0 + A_k q^2/2$ can model very general structures of the type $\bar{P}(Q) = \rmi f(q_0) + g(q^2)$, with arbitrary functions $f$ and $g$. A disadvantage of our choice of regulators is that we do not regularize frequencies and, thus, all frequencies $q_0$ contribute to the flow at each scale $k$. It is one of the purposes of this work to estimate the resulting error of this regularization scheme.

We can rescale the inverse boson propagator in Eq.~\eqref{appA-5} by $A_k^{-1}$, such that the gradient term is given by $q^2/2$ for each $k$. This is achieved by expressing the effective average action in terms of the renormalized field $\phi = \phi_k = A_k^{1/2} \bar{\vphi}$. The corresponding renormalized propagator reads
\begin{align}
 \label{appA-6} P_{\phi,k}(Q) = \frac{1}{A_k} \bar{P}_{\phi,k}(Q) = \rmi S_k q_0 + \frac{1}{2} q^2,
\end{align}
with $S_k=Z_k/A_k$. Due to this modification, the gradient coefficient $A_k$ drops out of the flow equations. The flow of $A_k$ enters the flow of renormalized vertices and propagators only through the anomalous dimension
\begin{align}
 \label{appA-7} \eta_k = - \frac{1}{A_k} k \partial_k A_k.
\end{align}
Whereas $Z_k$ and $A_k$ show a strong running with $k$, the quantities $S_k$ and $\eta_k$ are of order unity.

So far, we did not specify the form of the effective potential $\bar{U}_k(\bar{\rho},\mu)$. In principle, its form may be resolved by solving a partial differential equation in the variables $k$ and $\bar{\rho}$, which is obtained from the flow equation of $\Gamma_k$ for constant boson mean field. For the second order phase transitions expected here, it is qualitatively sufficient to restrict to a $\bar{\vphi}^4$-approximation of $\bar{U}$. More explicitly, we write in terms of the renormalized field $\rho = \phi^*\phi$:
\begin{align}
 \nonumber U_k(\rho) = \bar{U}_k(\bar{\rho}) = &\mbox{ }m^2_{\phi,k} (\rho-\rho_{0,k}) +\frac{\lambda_{\phi,k}}{2}(\rho -\rho_{0,k})^2 \\
  \label{appA-8}&- n_k \delta \mu + \alpha_k (\rho-\rho_{0,k})\delta \mu.
\end{align}
The minimum of $U_k$ is a $k$-dependent quantity. This allows to distinguish symmetric (ordered) and disordered regimes of the flow via $\rho_{0,k}=0$ and $\rho_{0,k}>0$, respectively. We have $m^2_{\phi,k}>0$ in the former and $m^2_{\phi,k}=0$ in the latter case. At zero temperature, the system always ends up in the superfluid phase for $k=0$, hence $\rho_{0,k=0}>0$ and $m^2_{\phi, k=0}=0$. The superfluid gap in our truncation is given by
\begin{align}
 \label{appA-9} \Delta_k = \sqrt{h_k^2 \rho_{0,k}},
\end{align}
with renormalized Feshbach coupling $h_k^2=h^2_\Lambda/A_k$.

The effective potential depends on the chemical potential $\mu$. It is related to the pressure $P$ according to
\begin{align}
 \label{appA-10} P(\mu) = - U_{k}(\rho_{0,k},\mu)\Bigr|_{k=0}.
\end{align}
The density $n=\partial P/\partial \mu$ can be computed by solving the flow equation for $U_k$ for two infinitesimally close values of the chemical potential and a subsequent finite difference according to $n=\Delta P/\Delta \mu$. Here we work with a flowing density which satisfies $n_{k=0}=n$. Moreover, we  approximate $\partial_\mu U_k' \approx \partial_\mu m^2_{\phi,k}=\alpha_k$. The flow equations for $n_k$ and $\alpha_k$ are obtained from the flow of the effective potential by virtue of a derivative with respect to the offset $\delta \mu$. The latter measures a difference from the actual chemical potential $\mu$ which also enters the fermion cutoff. In the same spirit we may write $\delta \rho = \rho-\rho_{0,k}$ in Eq.~\eqref{appA-8}.

The initial conditions for the running couplings $\eta_k$, $S_k$, $h^2_k$, $m^2_{\phi,k}$, $\lambda_{\phi,k}$, $\rho_{0,k}$, $n_k$, and $\alpha_k$ introduced here have to be chosen such that the boundary condition $\Gamma_\Lambda=S$ of the flow equations is satisfied. We show below that only the initial condition for the boson detuning $\nu_\Lambda$ constitutes a relevant parameter of the system. All other quantities are attracted to fixed point values in the early stages of the flow, i.e. for large $k$. For this reason, their initial values are not important and we may start directly at the fixed point.

\section{Flow equations}\label{append_flow_equations}

The flow equations for the running couplings defined in App.~\ref{append_truncation} are obtained from the flow equation for $\Gamma_k$ by appropriate projection prescriptions. The resulting set of coupled ordinary differential equations in $k$ can be solved by means of standard numerical techniques.

The flow equation for the effective potential $U_k(\rho)$ is found for a constant bosonic background field $\phi=\sqrt{2\rho}$ and vanishing fermion mean field. We have
\begin{align}
 \label{appB-1} k \partial_kU_k(\rho) = \eta_k \rho U_k'(\rho) + \dot{U}_k^{(B)}(\rho)+\dot{U}_k^{(F)}(\rho).
\end{align}
Herein, the first term takes into account that the effective average potential has a trivial running due to the fact that its argument $\rho=A_k\bar{\rho}$ is scale-dependent.  A prime denotes a derivative with respect to the variable $\rho$. The second and third term, respectively, constitute the contribution from bosonic and fermionic fluctuations to the flow. Within our truncation, these contributions to the flow generator $\zeta(\rho)$ for $U(\rho)$ read
\begin{align}
 \label{appB-2} \dot{U}_k^{(B)}(\rho) &= \frac{\sqrt{2}k^5}{3\pi^2 S_k} \Bigl(1-\frac{\eta_k}{5}\Bigr) \frac{1+(w_1+w_2)/2}{\sqrt{(1+w_1)(1+w_2)}},\\
 \label{appB-3} \dot{U}_k^{(F)}(\rho) &= - \frac{k^5}{3\pi^2 \sqrt{1+w_3}}\Bigl(\ell_1(\tilde{\mu})-\frac{w_3}{1+w_3}\ell_2(\tilde{\mu})\frac{\delta \mu}{k^2}\Bigr),
\end{align}
with $\ell_a(x)=\theta(x+1)(x+1)^{3/2}-(1+(-1)^a)\theta(x)x^{3/2}+(-1)^a\theta(x-1)(x-1)^{3/2}$ and Heaviside step function $\theta(x)$. We introduced the generalized dimensionless ``masses''
\begin{align}
 \label{appB-4} w_1 &= \frac{U_k'(\rho)}{k^2},\ w_2 = \frac{U_k'(\rho)+2\rho U_k''(\rho)}{k^2},\\
 \label{appB-5} w_3 &= \frac{h_k^2 \rho}{k^4},\ \tilde{\mu} = \frac{\mu}{k^2}.
\end{align}
These ratios control the decoupling of modes as $k$ is evolved towards zero. Taking derivatives of Eq.~\eqref{appB-1} with respect to $\rho$ and $\delta\mu$, we obtain the flow equations for $m^2_{\phi,k}$, $\lambda_{\phi,k}$, $n_k$ and $\alpha_k$. 

For the derivation of the flow equations for the wave function renormalization $Z_k$ and gradient coefficient $A_k$ we first consider the inverse boson propagator in the $(\phi_1,\phi_2)$-basis. Within our truncation it is given by
\begin{align}
 \label{appB-6} \bar{G}^{-1}_{\phi}(P) = \begin{pmatrix} A_k \left( \frac{p^2}{2}   + U_k' + 2 \rho U''\right) & - Z_k p_0 \\ Z_k p_0 & A_k \left( \frac{p^2}{2}   + U_k' \right) \end{pmatrix}.
\end{align}
We observe that the flow equations for $Z_k$ and $A_k$ can be obtained from the $12$- and $22$-components of the flow equation for $G^{-1}_\phi$, respectively. (Beyond our truncation the $11$-component contains an additional term $\sim A_k'(\rho_0)$ arising from a possible field dependence of $A_k(\rho)$, which is not present in the $22$-component. Therefore, the latter is more suited for the projection of $A_k$.) We have
\begin{align}
 \label{appB-7} \partial_k \bar{G}^{-1}_{\phi,ij}(P) \delta(P+P') = \frac{\delta^2\partial_k \Gamma_k}{\delta \bar{\phi}_i(P')\delta\bar{\phi}_j(P)}\Bigr|_{\rho_{0,k}}.
\end{align}

We define the finite difference projection
\begin{align}
 \label{appB-8} Z_k &= -\frac{\bar{G}^{-1}_{\phi,12}(p_0,0)-\bar{G}^{-1}_{\phi,12}(0,0)}{p_0}\Bigr|_{p_0=\cZ^2k^2},\\
 \label{appB-9} A_k &= 2 \frac{\bar{G}^{-1}_{\phi,22}(0,p^2)-\bar{G}^{-1}_{\phi,22}(0,0)}{p^2}\Bigr|_{p=c_{{\scriptscriptstyle A}} k}.
\end{align}
For $c_{{\scriptscriptstyle Z}}, c_{{\scriptscriptstyle A}} \to 0$ this becomes a differentiation at $P=0$. To relate the finite differences to the flow equation \eqref{appB-7}, we compute
\begin{align}
 &\nonumber \partial_k A_k =2 \partial_k \frac{\bar{G}_{\phi,22}^{-1}(0,c_{{\scriptscriptstyle A}}^2k^2)-\bar{G}_{\phi,22}^{-1}(0,0)}{c_{{\scriptscriptstyle A}}^2k^2}\\
 \label{appB-10} &=2 \frac{\partial_k\bar{G}_{\phi,22}^{-1}(0,c_{{\scriptscriptstyle A}}^2k^2)-\partial_k\bar{G}_{\phi,22}^{-1}(0,0)}{c_{{\scriptscriptstyle A}}^2k^2}  \\
 \nonumber &+4 \Biggl( \frac{\partial \bar{G}_{\phi,22}^{-1}}{\partial p^2}(0,c_{{\scriptscriptstyle A}}^2k^2)-\frac{\bar{G}_{\phi,22}^{-1}(0,c_{{\scriptscriptstyle A}}^2k^2)-\bar{G}_{\phi,22}^{-1}(0,0)}{c_{{\scriptscriptstyle A}}^2k^2}\Biggr).
\end{align}
The second term in brackets is in general non-zero, but vanishes within our truncation. Thus, the flow equation of the finite difference is given by the finite difference of the flow equations. The same holds for $\partial_k Z_k$ with a similar derivation.

The flow equation for $S_k=Z_k/A_k$ is evaluated at vanishing external momentum such that the momentum integral can be carried out analytically. Since our choice of regulators is frequency independent, the Matsubara summation can also be performed explicitly. We find
\begin{align}
 \label{appB-11} k \partial_k S_k = \eta_k S_k + \dot{S}_k^{(B)} + \dot{S}^{(F)}_k,
\end{align}
with
\begin{align}
 \nonumber \dot{S}_k^{(B)} &= \frac{2\sqrt{2}\lambda_{\phi,k}^2\rho_{0,k}}{3\pi^2 k(1+w_2)^{3/2}}\Bigl(1-\frac{\eta_k}{5}\Bigr)\\
 \nonumber &\times \frac{-16+2(c_{{\scriptscriptstyle Z}}^4 S_k^2-4)w_2+(28+c_{{\scriptscriptstyle Z}}^4S_k^2)w_2^2/2+6 w_2^3}{(4+c_{{\scriptscriptstyle Z}}^4 S_k^2+4w_2)^2},\\
 \nonumber \dot{S}_k^{(F)} &= \frac{h_k^2}{3 \pi^2 k(1+w_3)^{3/2}}\ell_2(\tilde{\mu})\\
 \label{appB-12}&\times \frac{-4(1+w_3)(2-w_3)+c_{{\scriptscriptstyle Z}}^4 w_3}{(c_{{\scriptscriptstyle Z}}^4+4(1+w_3))^2}.
\end{align}
The function $\ell_2(x)$ has been defined below Eq.~\eqref{appB-3}. In the limit $c_{{\scriptscriptstyle Z}}\to 0$ we recover the results from Ref.~\cite{PhysRevA.76.021602, PhysRevA.76.053627, ANDP:ANDP201010458}. 

The flow equation of $A_k$ for general $\cA$ is given in terms of a two-dimensional momentum integral, which has to be evaluated numerically at each step of the integration of the flow equations. We have $\eta_k(\cA)=\eta_k^{(B)}(\cA)+\eta_k^{(F)}(\cA)$ with
\begin{align}
 \nonumber \eta_k^{(B)} = &\frac{8\rho_0 \lambda_{\phi}^2}{p^2} \int_Q \frac{\dot{R}_\phi}{A_k} \Biggl\{  \frac{L_\phi (\vec{q}+\vec{p})-L_\phi (\vec{q})}{\mbox{det}_B(\vec{q})\mbox{det}_B(\vec{q}+\vec{p})}\Bigl(1-\frac{2(L_\phi (\vec{q}))^2}{\mbox{det}_B(\vec{q})}\Bigr)\\
 \label{ad1} &+\frac{L_\phi (\vec{q})}{\mbox{det}_B(\vec{q})}\Bigl(\frac{1}{\mbox{det}_B(\vec{q})}-\frac{1}{\mbox{det}_B(\vec{q}+\vec{p})}\Bigr)\Bigg\}_{p^2=\cA^2 k^2},\\
 \nonumber \eta_k^{(F)}= &\frac{4h^2}{p^2} \int_Q \dot{R}_\psi \Biggl\{ \frac{L_\psi (\vec{q}+\vec{p})-L_\psi (\vec{q})}{\mbox{det}_F(\vec{q})\mbox{det}_F(\vec{q}+\vec{p})}\Bigl(1-\frac{2(L_\psi (\vec{q}))^2}{\mbox{det}_F(\vec{q})}\Bigr)\\
 \label{ad2} &+\frac{L_\psi (\vec{q})}{\mbox{det}_F(\vec{q})}\Bigl(\frac{1}{\mbox{det}_F(\vec{q})}-\frac{1}{\mbox{det}_F(\vec{q}+\vec{p})}\Bigr)\Biggr\}_{p^2=\cA^2 k^2},
\end{align}
for the bosonic and fermionic contributions to the anomalous dimension, respectively. We introduced $\dot{R}_{\phi/\psi} = k \partial_k R_{\phi/\psi}(\vec{q}^2)$ and
\begin{align}
 L_\phi (\vec{q}) &= \frac{1}{2} \vec{q}^2 + R_\phi(\vec{q}^2) + m^2_\phi + \rho_0 \lambda_{\phi},\\
 \mbox{det}_B(\vec{q}) &= S_k^2 q_0^2 +(L_\phi (\vec{q}))^2-(\rho_0\lambda_{\phi})^2,\\
 L_\psi (\vec{q}) &= \vec{q}^2-\mu+R_\psi(\vec{q}^2),\\
 \mbox{det}_F(\vec{q}) &= q_0^2 + (L_\psi (\vec{q}))^2 +h^2\rho_0.
\end{align}
The Matsubara summations in Eqs.~(\ref{ad1}) and (\ref{ad2}) can again be evaluated analytically. Since 
\begin{align}
 \frac{\dot{R}_\phi}{A_k} = \Bigl(2k^2-\eta_k(k^2-q^2/2)\Bigr)\theta(k^2-q^2/2)
\end{align}
 depends linearly on $\eta_k$, we have to solve the linear system of Eqs.~(\ref{ad1})-(\ref{ad2}) to obtain a closed expression for $\eta_k(\cA)$. The formula for $\eta_k(\cA=0)$ is given in Refs.~\cite{ANDP:ANDP201010458}.

The flow equation for the Tan contact has been derived in Ref.~\cite{2012arXiv1209.5641B}. At zero temperature it is given by 
\begin{align}
  \nonumber k \partial_k C_k 
  = &
    \frac{h_k^2}{4}(\eta_k \rho_0 + k \partial_k \rho_0) \\
  &  - \frac{h_k^2 k^3 w_2^2}{24 \sqrt{2} \pi^2 S_k (1+w_2)^{3/2}}
  \Bigl(1-\frac{\eta_k}{5}\Bigr).
\end{align}

\section{Universality}\label{universality}
The running couplings of our truncation have to be equipped with appropriate initial conditions which ensure that the flow satisfies $\Gamma_\Lambda=S$. We show here that, for a broad Feshbach resonance (as it is realized with $^6$Li atoms), only the bosonic detuning term $\nu_\Lambda$ constitutes a relevant parameter, whereas all other couplings lose memory of their initial values. This universal behavior reflects the fact that for a broad Feshbach resonance the microscopic model is completely characterized by the value of the scattering length.

In the early stages of the renormalization group flow, i.e. for $k \lesssim \Lambda$, we have $\mu \ll k^2$ and $\rho_{0,k}=0$. Therefore, we can set $w_2=w_1$, $w_3 = \tilde{\mu}=0$ in the flow equations. The chemical potential does not influence the flow for large $k$, because the high momentum fluctuations cannot resolve its particular value. Hence, the  many-body system evolves similar to the vacuum system in this regime.

The flow equation for the dimensionless renormalized Feshbach coupling $\tilde{h}^2_k=h^2_k/k$ is given by
\begin{align}
 \label{appC-1} k \partial_k \tilde{h}^2_k = (\eta_k -1) \tilde{h}^2_k
\end{align}
for large $k$. The anomalous dimension in this regime is given by $\eta_k \sim \tilde{h}^2_k$ with a positive prefactor. Hence, if we start with a sufficiently large value for $\tilde{h}^2_\Lambda=h^2_\Lambda/\Lambda$, we have $\partial_k \tilde{h}^2_k >0$. This reduces the value of $\tilde{h}^2_k$ until we have $\eta_k =1$. At this point, the flow of the dimensionless Feshbach coupling stops and $h^2_k/k$ remains constant until $k$ becomes of the size of the chemical potential or the dimer binding energy (on the BEC side of the crossover). The intermediate regime of the flow, where the anomalous dimension is given by its fixed point value
\begin{align}
 \label{appC-2} \eta_\star =1,
\end{align}
is called the universal regime. In the subsequent paragraphs we assume $k$ to be in this range.

The fine-tuning of the bosonic mass term $m^2_{\phi,\Lambda}= U_{\Lambda}'(0)= \nu_\Lambda -2\mu$ has been discussed in detail in Ref.~\cite{ANDP:ANDP201010458}. For every choice of $a$, there exists a unique initial value $\nu_\Lambda=\nu_\Lambda(a)$, which describes a system with fermion scattering length $a$. We do not repeat the derivation of this result here, because the formulas are not altered by our finite difference projection with $c_{{\scriptscriptstyle Z}}, c_{{\scriptscriptstyle A}} \neq 0$. 

In contrast to the detuning from resonance, we find a modified fixed point structure for the flow of $h^2_k$, $S_k$, $\lambda_{\phi,k}$ and $\alpha_k$ in the early stages of the flow due to the more general projection prescription for $A_k$ and $Z_k$. The fixed point of $\tilde{h}^2_k$ is determined by Eq.~\eqref{appC-2} with
\begin{align}
 \label{appC-3} \eta_k = \frac{h^2_k}{6 \pi^2 k} x_\eta.
\end{align}
The positive constant $x_\eta$ depends on the choice of the specification parameter $c_{{\scriptscriptstyle A}}$. We find numerically
\begin{align}
 \label{appC-4} 0 \leq x_\eta(c_{{\scriptscriptstyle A}}) \leq 1 \text{ for all } c_{{\scriptscriptstyle A}}.
\end{align}
For $c_{{\scriptscriptstyle A}}=0$ we have $x_\eta =1$. The scaling behavior of the Feshbach coupling in the universal regime is deduced from Eq.~\eqref{appC-3} to be
\begin{align}
 \label{appC-5} h^2_\star = \frac{6 \pi^2 k}{x_\eta}.
\end{align}
Assuming $h^2_k$ to be at its fixed point value, we find
\begin{align}
 \label{appC-6} k \partial_k S_k &= \eta_\star S_k - \frac{h^2_\star}{6 \pi^2 k}x_S = S_k - \frac{x_S}{x_\eta},\\
 \label{appC-7} k \partial_k \alpha_k &= \eta_\star \alpha_k + \frac{h^2_\star}{3 \pi^2 k} = \alpha_k + \frac{2}{x_\eta}
\end{align}
with fixed point solutions
\begin{align}
 \label{appC-8} S_\star = x_S/x_\eta,\\
 \label{appC-9} \alpha_\star = -2/x_\eta.
\end{align}
Both fixed points correspond to stable solutions. From the analytical result for the flow of $S_k$ given in Eq.~\eqref{appB-12} we read off
\begin{align}
 \label{appC-10} x_S = \frac{1}{(1+c_{{\scriptscriptstyle Z}}^4/4)^2}.
\end{align}
We have $0 \leq x_S \leq 1$, with $x_S=1$ for $c_{{\scriptscriptstyle Z}}=0$. We conclude that choosing $c_{{\scriptscriptstyle A}}, c_{{\scriptscriptstyle Z}}\neq 0$ modifies the universal values of $S_k$ and $\alpha_k$. 

The flow of the dimer-dimer interaction strength $\lambda_{\phi,k}$ for large $k$ receives contributions from both fermionic and bosonic diagrams. With $m^2_{\phi,\star}=k^2$, the corresponding flow equation
\begin{align}
 \label{appC-11} k \partial_k \lambda_{\phi,k} = 2 \eta_\star \lambda_{\phi,k} -\frac{h^4_\star}{4\pi^2 k^3}- \frac{\sqrt{2}k^5(1-\eta_\star/5)}{3 \pi^2 (k^2+m^2_{\phi,\star})^2S_\star}\lambda_{\phi,k}^2
\end{align}
has a fixed point solution $\tilde{\lambda}_{\phi,\star}=\tilde{\lambda}_{\phi,\star}(x_\eta,x_S)$ found from the quadratic equation $\partial_k \tilde{\lambda}_{\phi,k}=\partial_k (k\lambda_{\phi,k})=0$.

We have seen that, due to the existence of fixed point solutions of the flow, the running couplings rapidly approach their universal values. In particular, the universal regime is entered far before the chemical potential enters the renormalization group flow as a scale. Therefore, except for the mass term, the initial values of the couplings at scale $k=\Lambda$ are irrelevant. In particular, even for initially non-dynamical bosonic degrees of freedom with $Z_\Lambda=A_\Lambda=0$, the corresponding terms in the propagator are immediately generated. For this reason, we may start with $Z_\Lambda,A_\Lambda \neq 0$ as well. To summarize, we choose
\begin{align}
 \nonumber h^2_\Lambda &= \frac{6\pi^2\Lambda}{x_\eta},\ m^2_{\phi,\Lambda}=\nu_\Lambda(a) - 2\mu,\ \lambda_{\phi,\Lambda} = \frac{\tilde{\lambda}_{\phi,\star}}{\Lambda},\\
 \label{appC-12} S_\Lambda &= x_S/x_\eta,\ n_\Lambda = \theta(\mu) \frac{\mu^{3/2}}{3\pi^2},\ \alpha_\Lambda=-2/x_\eta.
\end{align}
The initial value of the flowing density results from the form of the microscopic fermion propagator.

\end{appendix}


\bibliographystyle{apsrev4-1}
\bibliography{references_error_estimates.bib}

\end{document}